\documentclass[aps,prl,reprint,showpacs,amsmath,amssymb,superscriptaddress,groupedaddress]{revtex4-1}
\usepackage{graphicx}
\usepackage{dcolumn}
\usepackage{epstopdf}
\usepackage{bm}
\usepackage{color}
\begin{document}
\title{Intrinsic Charge Carrier Mobility in Single-Layer Black Phosphorus}

\author{A.~N. Rudenko}
\email[]{a.rudenko@science.ru.nl}
\author{S. Brener}
\author{M.~I. Katsnelson}
\affiliation{\mbox{Institute for Molecules and Materials, Radboud University, Heijendaalseweg 135, 6525 AJ Nijmegen, Netherlands}}
\date{\today}

\begin{abstract}
We present a theory for single- and two-phonon charge carrier scattering in 
anisotropic two-dimensional semiconductors applied to single-layer black 
phosphorus (BP). We show that in 
contrast to graphene, where two-phonon processes due to the scattering by 
flexural phonons dominate at any practically relevant temperatures and are 
independent of the carrier concentration $n$, two-phonon scattering in BP is less important and can
be considered negligible at $n\gtrsim10^{13}$ cm$^{-2}$. At smaller $n$, however, phonons enter in the
essentially anharmonic regime.
Compared to the hole mobility, which does not exhibit strong 
anisotropy between the principal directions of BP 
($\mu_{xx}/\mu_{yy}\sim1.4$ at $n=10^{13}$ cm$^{-2}$ and $T=300$ K), the electron mobility is found to be significantly 
more anisotropic ($\mu_{xx}/\mu_{yy}\sim6.2$). Absolute values of $\mu_{xx}$ do not exceed 250 (700) cm$^2$V$^{-1}$s$^{-1}$ 
for holes (electrons), which can be considered as an upper limit for the mobility in BP at room temperature.
\end{abstract}


\maketitle

Electron-phonon scattering is considered to be the main factor limiting intrinsic charge-carrier mobility in 
graphene \cite{Katsnelson-Book,Stauber2007,Morozov2008,Mariani2008,Mariani2010,Castro2010b,Ochoa2011}.
Flexural phonons (out-of-plane vibrations) are especially 
important in this respect because they provide the dominant contribution to the resistivity at room temperature \cite{Castro2010b,Ochoa2011}.
Recently, many new two-dimensional (2D) materials have attracted attention \cite{Geim2013}, such as
hexagonal boron nitride \cite{Xu2013}, stoichiometric graphene derivatives \cite{Elias2009,Nair2010},
transition-metal dichalcogenides \cite{Wang2012,Xu2013}, and black phosphorus (BP) \cite{Ling2015}. All these materials
are typically more defective than graphene and are characterized by significantly smaller electron mobility;
therefore, much less is known experimentally on their intrinsic transport properties \cite{Baugher2013,Yu2014,Li2014,Liu2014,Koenig2014,Buscema2014,Doganov2014,Schmidt2015}.

Comprehensive theories have been developed to describe the mechanism of phonon scattering in graphene \cite{Ochoa2011}.
The application of those is, however, not straightforward to systems with reduced symmetries that give rise to anisotropy of electronic and vibrational properties.
At the same time, anisotropy of 2D materials in not uncommon. It can naturally arise in finite-size samples 
and be governed by the shape (e.g., nanoribbons) \cite{Xu2009} or can be determined by external conditions such as defects \cite{Heine2013}, mechanical 
strain \cite{Pereira2009}, or contact potentials \cite{Park2008}. Few-layer black phosphorus is the most prominent example among 2D materials with inherent anisotropy \cite{Ling2015}.
Early attempts to describe intrinsic mobility in ultrathin BP were based on isotropic transport theory and were focused on single-phonon processes only \cite{Qiao2014,Liao2015}.

In this Letter, we develop a theory for phonon-limited transport in anisotropic 2D semiconductors. 
We obtain general expressions for the scattering matrix of single- and two-phonon processes, where both in-plane and flexural acoustic phonons
are included. The theory is applied to monolayer black phosphorus, for which the relevant parameters are estimated from first principles.

For isotropic materials, phonon limited dc conductivity is usually calculated using the standard semiclassical
Boltzmann theory \cite{Ziman2001}. The solution of the anisotropic Boltzmann equation is more involved
even within the simple effective mass approximation \cite{Tokura,LiuLow}. Here, we use an alternative approach, namely,
the Kubo-Nakano-Mori method, which does not rely on any symmetry constraints; it is equivalent to the variational solution 
of the Boltzmann equation in the isotropic limit at $k_BT\ll \varepsilon_F$ \cite{Katsnelson-Book}, where $k_BT$ is 
the temperature in energy units and $\varepsilon_F$ is the Fermi energy. The formula for the $x$ component of the conductivity 
reads \cite{Ziman2001}
\begin{equation}
\sigma_{xx}=\frac{e^2}{2S}\sum_{\bf k} \tau_{xx} v^{x2}_{{\bf k}} \left( -\frac{\partial f}{\partial \varepsilon_{\bf k}} \right),
\label{sigmax}
\end{equation}
where $e$ is the elementary charge, $S$ is the sample area, $v^x_{{\bf k}}$ is the $x$ component of the carrier velocity, $f=\{1+\mathrm{exp}[(\varepsilon_{\bf k}-\varepsilon_F)/k_BT]\}^{-1}$
is the Fermi-Dirac distribution function, $\varepsilon_{\bf k}$ is the carrier energy,
and $\tau_{xx}$ can be understood as the scattering relaxation time of carriers in the $x$ direction. In Eq.~(\ref{sigmax}) we assume that cross terms such as $\tau_{xy}$ can be eliminated for
symmetry reasons. 
In turn, the expression for $\tau_{xx}$ has the form \cite{Katsnelson-Book}

\begin{equation}
\frac{1}{\tau_{xx}}=\frac{\pi}{\hbar}\frac{\sum_{{\bf k}{\bf k}'}\delta(\varepsilon_{{\bf k}}-\varepsilon_{\bf k'}) \left( -\frac{\partial f}{\partial \varepsilon_{{\bf k}}} \right) (v^x_{{\bf k}}-v^x_{{\bf k}'})^2\langle |V^{\mathrm{eff}}_{{\bf k}{\bf k}'}|^2 \rangle}{\sum_{\bf k}v^{x2}_{\bf k}  \left( -\frac{\partial f}{\partial \varepsilon_{{\bf k}}} \right)},
\label{taux}
\end{equation}
where ${\bf k}={\bf k}'+{\bf q}$ with ${\bf q}$ being the phonon wave vector, $V^{\mathrm{eff}}_{{\bf k}{\bf k}'}$ is the scattering matrix, and $\langle ... \rangle $ denote ensemble 
averaging over the phonon states. We note that Eq.~(\ref{taux}) is derived assuming that carrier scattering
is elastic, i.e., the phonon energy satisfies $\hbar \omega({\bf q}\approx{\bf k}_F) \ll \varepsilon({\bf k}_F)$, where ${\bf k}_F$ is the
Fermi wave vector. Importantly, Eq.~(\ref{taux}) does not assume that the scattering is isotropic.

We now turn to the evaluation of electron-phonon scattering matrices, $\langle |V^{\mathrm{eff}}_{{\bf k}{\bf k}'}|^2 \rangle$. 
In principle, those can be accurately calculated from
first principles using well-established interpolation techniques \cite{Giustino}. Such numerical approaches, however, are very sensitive to the Brillouin zone (BZ) sampling and thus are hardly 
applicable to two-phonon processes, whose description involves additional BZ integration. Here, we restrict ourselves to the long-wavelength limit and derive 
$\langle |V^{\mathrm{eff}}_{{\bf k}{\bf k}'}|^2 \rangle$ analytically for both single- and two-phonon processes using the concept of deformation potentials. Apart from being general,
in certain cases our approach allows one to analytically calculate the scattering rates given by Eq.~(\ref{taux}). Similar approaches have been previously followed to describe single-phonon 
scattering in 3D materials \cite{Herring}.

For 2D materials with orthorhombic (and higher) symmetries, elastic energy associated with in-plane deformations in the harmonic approximation reads \cite{Landau}
\begin{equation}
\overline{E}=\frac{1}{2} \! \int \! d^2{\bf r} \left[ C_{11} u_{xx}^2 + C_{22} u_{yy}^2 + 2C_{12} u_{xx} u_{yy} + 4C_{66} u_{xy}^2 \right],
\label{energy_inplane}
\end{equation}
where $u_{xx}$, $u_{yy}$, and $u_{xy}$ are components of the strain tensor, and $C_{11}$, $C_{22}$, $C_{12}$, and $C_{66}$ are elastic constants. 
In the long-wavelength limit, the effective scattering potential of charge carriers induced by in-plane acoustic phonons $\overline{V}^{\mathrm{eff}}_{\bf q}$ can be written in terms of the diagonal
components of the deformation potential tensor $\overline{g}_{\alpha}$ ($\alpha=x,y$) as $\overline{V}^{\mathrm{eff}}_{\bf q} = \overline{g}_{\alpha} u_{q_{\alpha}} q_{\alpha}$ (summation over repeated Greek indices is implied throughout the Letter), where $u_{q_{\alpha}}$ is 
the $\alpha$ component of the reciprocal displacement vector $u_{\bf q}$.
Taking the square of $\overline{V}_{\bf q}^{\mathrm{eff}}$ and averaging over the phonon states, we get
\begin{equation}
\langle |\overline{V}^{\mathrm{eff}}_{{\bf q}}|^2 \rangle = \langle u^*_{q_{\alpha}}u_{q_{\beta}} \rangle \overline{g}_{\alpha} \overline{g}_{\beta} q_{\alpha} q_{\beta},
\label{veffbar}
\end{equation}
where $\langle u^*_{q_{\alpha}}u_{q_{\beta}} \rangle = k_BT (A^{\bf q})^{-1}_{\alpha\beta}$ is the temperature-dependent correlation function for in-plane fields with 
$A^{\bf q}_{\alpha \beta}$ being the reciprocal force-constant matrix \cite{SI}. Here, we utilized the fact
that phonons can be considered classically at moderate temperatures, i.e., $\hbar\omega({\bf q}\approx{\bf k}_F)\ll k_BT$, which always holds for $|{\bf k}_F| \ll a^{-1}$, where 
$a$ is the interatomic distance. 
The same criterion ensures that the effect of the wave function overlap between the initial and final states is negligible \cite{Overlap}.
The final expression for the scattering matrix associated with in-plane phonons takes the form
\begin{widetext}
\begin{equation}
\langle |\overline{V}^{\mathrm{eff}}_{{\bf q}}|^2 \rangle = k_BT \frac{C_{66}(\overline{g}_x^2q_x^4+\overline{g}_y^2q_y^4-2\overline{g}_x\overline{g}_yq_x^2q_y^2) + (C_{22}\overline{g}_x^2+C_{11}\overline{g}_y^2-2C_{12}\overline{g}_x\overline{g}_y) q^2_xq^2_y} {C_{66}(C_{11}q_x^4+C_{22}q_y^4-2C_{12}q_x^2q_y^2)+(C_{11}C_{22}-C_{12}^2)q_x^2q_y^2}.
\label{Veff_inplane}
\end{equation}
\end{widetext}
Unlike the isotropic case ($C_{11}=C_{22}$), where $\langle |\overline{V}^{\mathrm{eff}}_{{\bf q}}|^2 \rangle=k_BT\overline{g}^2/C_{11}$ does not depend on ${\bf q}$, the scattering
probability of anisotropic systems displays a sophisticated {\bf q} dependence.

Let us now consider scattering on flexural phonons. Here, we focus on two-phonon processes only because single-phonon scattering involving flexural modes in BP is symmetry forbidden. 
Indeed, the point group ($D_{2h}$) corresponding to the space group of BP ($D_{2h}^7$) contains horizontal mirror plane symmetry operation ($\sigma_{h}$) \cite{Soares}, meaning that the scattering
potential $\overline{V}^{\mathrm{eff}}_{{\bf q}}$ is odd with respect to flexural deformations. This ensures that the corresponding matrix element 
$\langle |\overline{V}^{\mathrm{eff}}_{{\bf q}}|^2 \rangle$ vanishes \cite{Fischetti}.

In the presence of pure flexural deformations, elastic energy of an anisotropic membrane can be written as,
\begin{equation}
\widetilde{E}=\frac{1}{2} \! \int \! d^2{\bf r}\left[ \kappa_x (\partial^2_x h)^2 + \kappa_y (\partial^2_y h)^2 + 2\kappa_{xy} \partial^2_x h \partial^2_y h \right],
\label{energy_flex}
\end{equation}
where $h=h(x,y)$ is a field of out-of-plane displacements, and $\kappa_{x}$, $\kappa_{y}$, and $\kappa_{xy}$ are constants determining the flexural rigidity of the membrane. In this case,
the scattering potential is given by $\widetilde{V}^{\mathrm{eff}}_{{\bf q}}=\widetilde{g}_{\alpha\beta} f_{\alpha\beta}({\bf q})$, where 
$f_{\alpha\beta}({\bf q})=-\sum_{{\bf k}_1}k_{1\alpha}(q_{\beta}-k_{1\beta})h_{{\bf k}_1}h_{{\bf q}-{\bf k}_1}$ is the Fourier component corresponding to the tensor of flexural deformations 
$f_{\alpha\beta}({\bf r})=[\partial h({\bf r}) / \partial x_{\alpha}] [\partial h({\bf r}) / \partial x_{\beta}]$. 
After straightforward manipulations \cite{SI}, the scattering probability due to acoustic flexural phonons takes the form

\begin{eqnarray}
\langle |\widetilde{V}^{\mathrm{eff}}_{{\bf q}}|^2 \rangle=\sum_{{\bf p}}\left[ \widetilde{g}_x p_x(q_x-p_x)+\widetilde{g}_y p_y(q_y-p_y) \right ]^2 \nonumber \\
\times \langle h_{{\bf p}} h_{-{\bf p}}\rangle \langle h_{{\bf p}-{\bf q}} h_{{\bf q} - {\bf p}} \rangle, \quad \quad \quad \quad
\label{Veff_integral}
\end{eqnarray}
where the correlation function can be expressed as \cite{Katsnelson-Book}
\begin{equation}
\langle h_{{\bf q}} h_{-{\bf q}}\rangle = \frac{k_BT}{({\sqrt \kappa_x }q_x^2+{\sqrt \kappa_y} q_y^2)^2}
\label{corr_func}
\end{equation}
assuming that $\kappa_{xy}=\sqrt{\kappa_x \kappa_y}$.
Substituting Eq.~(\ref{corr_func}) into Eq.~(\ref{Veff_integral}) and evaluating the integral over ${{\bf p}}$ with logarithmic precision one finds \cite{SI}

\begin{align}
\langle |\widetilde{V}^{\mathrm{eff}}_{{\bf q}}|^2 \rangle &=\frac{k_B^2T^2\widetilde{g}_x^2}{4\pi r^{1/2} \kappa_x^2}\left[ \frac{1}{q_x^2+rq_y^2} \right] \left\{ \left[ 1 - \frac{8rq_x^2q_y^2}{(q_x^2+rq_y^2)} \right] \frac{(1-p)^2}{2} \right. \nonumber \\
&\left. + \frac{q_x^2-rq_y^2}{q_x^2+rq_y^2}\left( 1-p^2 \right) \mathrm{ln}\gamma + (1+p^2) \mathrm{ln}\gamma + \frac{(1+p)^2}{4}  
\vphantom{\frac{\sqrt{r}}{q_x^2+rq_y^2}} \right\},
\label{Veff_flex}
\end{align}
where we used the notation $r=(\kappa_y/\kappa_x)^{1/2}$ and $p= [(\widetilde{g}_y/\widetilde{g}_x) / (\kappa_y/\kappa_x)^{1/2}]$.
In Eq.~(\ref{Veff_flex}), $\gamma=\overline{k} / |{\bf q}^*|$, where $\overline{k}$ is a characteristic carrier wave vector ($\overline{k} \gg |{\bf q}^*|$) and 
${\bf q}^{*}$ is the critical wave vector, which determines the applicability of the harmonic approximation, i.e., independent carrier scattering on in-plane and flexural 
phonons \cite{cutoff,Nelson}. As for in-plane phonons, scattering matrix for flexural phonons exhibits a nontrivial ${\bf q}$ dependence compared to the isotropic case, where 
$\langle |\widetilde{V}^{\mathrm{eff}}_{{\bf q}}|^2 \rangle \sim k_B^2T^2/|{\bf q}|^2$.

In 2D materials with degenerate valleys like graphene and transition metal dichalcogenides, there is another important contribution to the scattering probability arising from the variation 
of hopping integrals upon deformation. This effect is equivalent to the presence of a gauge potential acting with opposite sign at each valley \cite{Vozmediano}, which shifts the position 
of valleys in BZ, leading to an additional scattering channel limiting mobility \cite{Castro2010b,Park,Sohier,ZLi,Rostami}. 
However, for single-valley systems like BP, such effects are forbidden as the position of the band edges in {\bf k}-space is protected by time-reversal 
symmetry. As can be explicitly shown within a tight-binding model \cite{Rudenko2014}, the variation of hopping integrals in BP is fully captured by the deformation potentials.

\begin{figure}[bp]
\includegraphics[width=0.50\textwidth, angle=0]{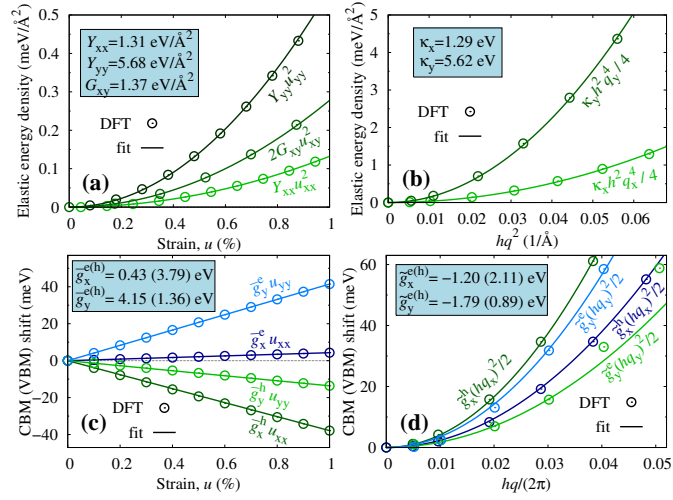}
\caption{(a),(b) Elastic energy and (c),(d) band-edge shifts 
as functions of in-plane and out-of-plane deformations in BP used 
to determine elastic constants and deformation potentials. 
CBM and VBM are the conduction band 
minimum and valence band maximum, respectively. Points correspond to DFT
calculations, whereas lines are the result of fitting with the constants shown
in the insets. In-plane deformations are induced by direction-dependent strain 
($u$), whereas out-of-plane deformations are characterized by
the wave vector $q$ and amplitude $h$ of a sinusoidal corrugation along the
armchair ($x$) and zigzag ($y$) directions \cite{SI}.}
\label{elastic}
\end{figure}

We now apply the presented theory for the calculation of direction-dependent carrier mobility in BP. We first determine parameters appearing in the expressions for the scattering 
matrix [Eqs.~(\ref{Veff_inplane}) and (\ref{Veff_flex})], i.e., elastic constants and deformation potentials, from first principles. In Fig.~\ref{elastic}, we show the elastic 
energy and band-edge shifts induced by in-plane and out-of-plane deformations calculated using density functional theory (DFT) \cite{SI}. Elastic constants can be obtained by fitting DFT energy
curves to the macroscopic expressions given by Eqs.~(\ref{energy_inplane}) and (\ref{energy_flex}) noting that $C_{11}=Y_{xx}(1-\nu_{xy}\nu_{yx})^{-1}$, $C_{22}=Y_{yy}(1-\nu_{xy}\nu_{yx})^{-1}$,
$C_{12}=C_{11}\nu_{xy}=C_{22}\nu_{yx}$, and $C_{66}=G_{xy}$. Here, $Y_{xx}$, $Y_{yy}$, and $G_{xy}$ are 2D Young moduli and shear modulus, respectively, whereas $\nu_{xy}=0.70$, $\nu_{yx}=0.16$ are Poisson 
ratios directly obtained from DFT calculations. The deformation potentials are determined similarly by fitting the corresponding DFT electron (hole) band shifts as shown in Figs.~\ref{elastic}(c) and \ref{elastic}(d). 
For symmetry reasons mentioned above, flexural deformations do not induce linear terms in the band shifts [see Fig.~\ref{elastic}(d)].

Dispersion $\varepsilon_{\bf k}$ of both electrons and holes in BP exhibits considerable deviations from the parabolic law, yielding energy-dependent density of states (DOS) as shown in Fig. 2. To 
capture the effects of nonparabolicity in the mobility calculations, we use the energy-dependent effective mass approximation \cite{SI}, which demonstrates good agreement with the results of 
first-principles $GW$ calculations (Fig.~\ref{dispersion}).

\begin{figure}[bp]
\includegraphics[width=0.50\textwidth, angle=0]{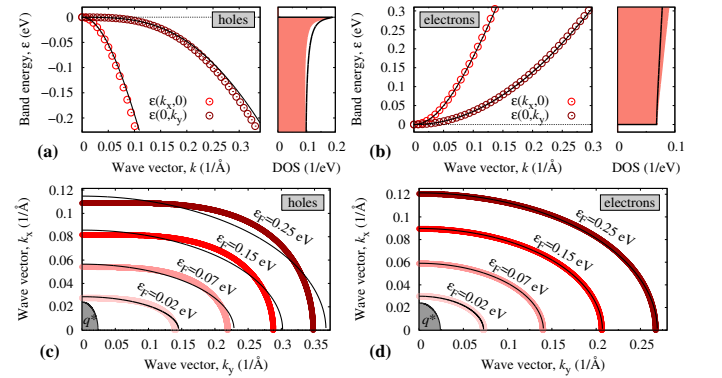}
\caption{(a),(b) Energy dispersion of electrons and holes
in BP along the armchair ($x$) and zigzag ($y$) directions with
the related DOS. (c),(d) Fermi contours 
$\varepsilon_F=\varepsilon(k_x,k_y)$ shown for the irreducible wedge of the BZ.
Points and thick lines are the result of $GW$ calculations
\cite{SI}, whereas thin black lines correspond to the model used in this Letter. Gray
area marks the phonon cutoff wave vector $q^*$ at $T=300$ K.}
\label{dispersion}
\end{figure}

The results on intrinsic mobility of electrons and holes in BP are presented in 
Fig.~\ref{mobility}. The total mobility in a specific direction is defined adopting Matthiessen's rule, $\mu^{-1}=\overline{\mu}^{-1}+\widetilde{\mu}^{-1}$,
where $\overline{\mu}=\overline{\sigma}/ne$ ($\widetilde{\mu}=\widetilde{\sigma}/ne$) is the corresponding contribution from single-phonon
(two-phonon) processes.
At $n\gtrsim10^{13}$ cm$^{-2}$, i.e., in the regime where the harmonic approximation is applicable [$\mathrm{ln}(\overline{k}/q^*)>1$],
single-phonon processes dominate for both electrons and holes at any practically relevant temperatures. 
This observation is in stark contrast with graphene, where two-phonon processes dominate independently of $n$ \cite{Castro2010b,Ochoa2011}.
Let us assume that the electron gas is degenerate ($\varepsilon_F/k_BT \gg 1$). The corresponding ratio of single-phonon $\overline{\mu}_{xx}$ and 
two-phonon $\widetilde{\mu}_{xx}$ mobilities then reads \cite{SI}
\begin{equation}
\frac{\overline{\mu}_{xx}}{\widetilde{\mu}_{xx}} = \left( \frac{\widetilde{g}_x}{\overline{g}_x} \right)^2 \left( \frac{\widetilde{A}_{xx}}{\overline{A}_{xx}} \right) \frac{C_{11}}{8\pi^2\kappa_x^2N(\varepsilon_F)} \mathrm{ln}(\sqrt{\mathrm{e}}\gamma) \frac{k_BT}{\varepsilon_F}, 
\label{ratio1}
\end{equation}
where $\mathrm{e}\approx2.718$ and $\overline{A}_{xx}$, $\widetilde{A}_{xx}$ are the cumulative anisotropic factors incorporating the effects of different kinds of anisotropies associated with single-phonon and two-phonon processes, respectively. One can see from Table \ref{parameters} that these factors are non-negligible and play an appreciable role in determining the mobility; this
is especially clear for electrons propagating in the armchair direction, where $\overline{A}^{e}_{xx} \approx 15$.

\begin{figure}[tbp]
\includegraphics[width=0.48\textwidth, angle=0]{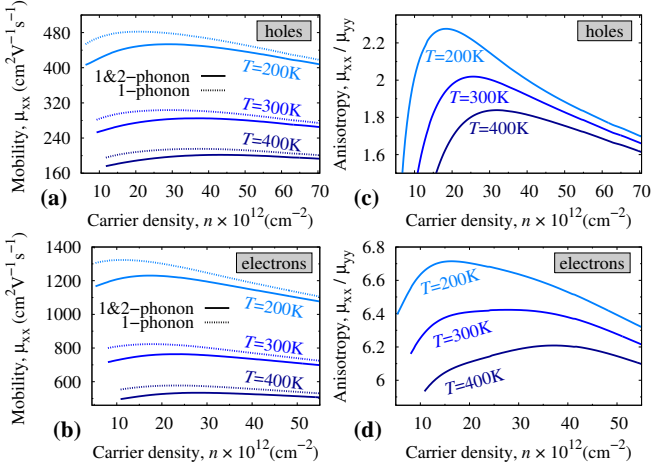}
\caption{(a),(b) Intrinsic carrier mobility ($\mu_{xx}$) of 
BP shown as a function of the carrier concentration ($n$) calculated 
along the armchair direction for different temperatures ($T$). (c),(d) 
Ratio between the 
mobilities along the armchair and zigzag directions ($\mu_{xx}/\mu_{yy}$) shown
for different $T$. Solid lines 
correspond to the contribution of both single-phonon and two-phonon scattering 
processes, whereas dashed lines correspond to single-phonon processes only. The lowest
depicted density corresponds to the regime with $\mathrm{ln}(\overline{k}/q^*)>1$.}
\label{mobility}
\end{figure}

    \begin{table}[bp]
    \centering
    \caption[Bset]{Single- and two-phonon contributions to carrier mobilities ($\overline{\mu}$ and $\widetilde{\mu}$) in BP calculated at \mbox{$n=5\times10^{13}$ cm$^{-2}$} and $T=300$ K along 
different transport directions (in cm$^2$V$^{-1}$s$^{-1}$) and the corresponding anisotropic factors ($\overline{A}$ and $\widetilde{A}$) indicating the role of anisotropy in each case. 
$\alpha$ is the prefactor of $k_BT/\varepsilon_F$ in Eq.~(\ref{ratio1}), determining the $\overline{\mu}/\widetilde{\mu}$ ratio.}
    \label{parameters}
 \begin{tabular}{cccccccccccc}
      \hline
 &  \multicolumn{2}{c}{holes} & & \multicolumn{2}{c}{electrons} & \\
\cline{2-3}
\cline{5-6}
                                &   \quad   Armchair  \quad    & \quad    Zigzag  \quad     &  & \quad    Armchair \quad     & \quad   Zigzag  \quad       &  \\
      \hline
     $\overline{\mu}$                &         292        &  157             &  &    738             &    114            &  \\
     $\widetilde{\mu}$              &    7$\times$10$^3$  &  8$\times$10$^3$ &  &  19$\times$10$^3$  &   16$\times$10$^3$ &  \\
     $\overline{A}$                  &    0.43            &  2.65            &  &   15.20           &   0.98            &  \\
     $\widetilde{A}$                &    0.43            &  3.86            &  &   0.58            &   1.18            &  \\
     $\alpha$                   &    0.89            &  0.37            &  &     0.99          &   0.17            &  \\
      \hline
    \end{tabular}
    \end{table}

In Eq.~(\ref{ratio1}), the numerical factor in front of $k_BT/\varepsilon_F$ (given in Table \ref{parameters} as $\alpha$ for $n=5\times10^{13}$ cm$^{-2}$) is less than unity for both electrons and 
holes in BP, meaning that single-phonon processes
always dominate for the degenerate electron gas, i.e., $\overline{\mu}\ll\widetilde{\mu}$. The only situation where two-phonon processes are expected to be significant corresponds to the case of a small doping ($n\ll10^{13}$ cm$^{-2}$).
In such a regime, however, anharmonic coupling between the phonons becomes significant ($\overline{k}/q^*<1$) in the charge carrier scattering, whose quantitative description requires a separate consideration.

It is instructive to make a comparison of single- and two-phonon scattering in BP with graphene. Given that $\overline{g}=\widetilde{g}$ for isotropic and atomically thin membranes, for 
equal carrier densities an estimate of the numerical factor in Eq.~\ref{ratio1} gives $\alpha \sim 27$ for graphene \cite{Remark}, which is
up to 2 orders of magnitude higher than in BP (see Table \ref{parameters}). Since DOSs are comparable at the given carrier concentration, the main difference stems from
the difference in elastic properties between the two materials. Indeed, $C/\kappa^2$ in graphene is 13 and 58 times larger than the corresponding ratio calculated for armchair and zigzag directions 
of BP, respectively. Physically, this is attributed to the combination of unusually high in-plane stiffness and superior flexibility of graphene compared to other 2D materials. At lower 
carrier densities, $\alpha$ in graphene becomes even larger because the energy-dependent DOS ensures that $\alpha \sim 1/\varepsilon_F$, favoring two-phonon scattering.

As can be seen from Figs.~\ref{mobility}(c) and \ref{mobility}(d), carrier mobilities in BP exhibit strongly different anisotropies $\mu_{xx}/\mu_{yy}$ for electrons and holes. At $n=10^{13}$ cm$^{-2}$
and $T=300$ K it amounts to $\sim$6.2 for electrons and $\sim$1.4 for holes. The latter value is in excellent agreement with the experimental value of $1.66$ obtained
in Ref.~[\onlinecite{Mishchenko}] for hole doping in few-layer BP. Within the approximations assumed above,
the ratio of the total mobility along different crystallographic directions is
\begin{equation}
\frac{\mu_{xx}}{\mu_{yy}} \approx \frac{m_y}{m_x}\frac{\overline{\tau}^{-1}_{yy}}{\overline{\tau}^{-1}_{xx}} = \frac{m_y}{m_x} \left( \frac{\overline{g}_y}{\overline{g}_x} \right)^2 \left( \frac{\overline{A}_{yy}}{\overline{A}_{xx}} \right) \left( \frac{C_{11}}{C_{22}} \right).
\end{equation}
The resulting anisotropy of the carrier mobility represents, therefore, a complex interplay between different anisotropic factors, among which are the anisotropies of effective masses, 
deformation potentials, and elastic constants. In BP, none of these factors can be considered as essentially isotropic, which makes nontrivial a simple qualitative description of
the observable anisotropic properties. Interestingly, at $n<10^{13}$ cm$^{-2}$ the two-phonon contribution favors the \emph{inverse} ($\mu_{xx}/\mu_{yy}<1$) anisotropy of the hole
mobility \cite{SI}. Such behavior is expected to be more pronounced at low carrier concentrations, and is indeed observed in recent scanning tunneling microscopy 
experiments \cite{Lu2015}.

The estimated absolute values of the mobilities along the armchair ($x$) direction at $n=10^{13}$ cm$^{-2}$ and $T=300$ K are $\sim$250 and $\sim$700 cm$^2$V$^{-1}$s$^{-1}$ for holes and electrons, 
respectively. The obtained values are rather low and can be considered as an upper limit for the mobility in BP at room temperature. At the higher and lower carrier concentrations there is
some decrease in mobility due to the energy dependence of the effective masses and the Fermi smearing effects, respectively.
Because single-phonon processes dominate, they cannot be easily suppressed by, e.g., encapsulation or depositing of BP samples on substrates.
For the same reason, at $n\gtrsim10^{13}$ cm$^{-2}$ the mobility is inversely proportional to the temperature ($\mu \sim T^{-1}$). 
Our estimate for the mobility values is found to be in a good agreement with available experimental data on field-effect mobilities in few-layer 
BP \cite{Li2014,Liu2014,Koenig2014,Buscema2014,Lu2015,Tayari2015}, suggesting that scattering on acoustic phonons is one of the main factors limiting the intrinsic mobility in BP.
We note, however, that at considerably higher temperatures ($T\gg300$ K) and higher carrier densities ($n\gg$10$^{13}$ cm$^{-2}$), other factors like optical phonons or 
electron-electron scattering (not considered here) might become more important.

To conclude, we have presented a consistent theory for the charge carrier scattering on acoustic phonons in anisotropic 2D systems. Both single-phonon and two-phonon processes are 
taken into account on equal footing.
The theory is applied to the calculation of intrinsic mobilities in single-layer black phosphorus, for which relevant parameters are obtained from first principles. We have
shown that, contrary to graphene, two-phonon processes governed by flexural phonons can be considered negligible at carrier concentrations $n\gtrsim10^{13}$ cm$^{-2}$. 
The estimated intrinsic mobility in
BP at $n\sim10^{13}$ cm$^{-2}$ and \mbox{$T=300$ K} do not exceed $\sim$250 and $\sim$700 cm$^2$V$^{-1}$s$^{-1}$ for holes and electrons.
Given that these values can be considerably reduced by other intrinsic and extrinsic scattering mechanisms, the application of BP in real devices might be hindered.

\begin{acknowledgments} 
The research has received funding from the 
European Union Seventh Framework Programme and Horizon 2020 Programme under Grants No.~604391 and No.~696656, Graphene Flagship, and from the Stichting voor Fundamenteel Onderzoek der Materie (FOM), which is financially supported by the Nederlandse Organisatie voor Wetenschappelijk Onderzoek (NWO).
\end{acknowledgments}

\maketitle

\onecolumngrid\newpage\twocolumngrid
\onecolumngrid
\begin{center}
\textbf{\large Supplemental Material: Intrinsic Charge Carrier Mobility \\ in Single-Layer Black Phosphorus}
\end{center}
\twocolumngrid

\setcounter{equation}{0}
\setcounter{figure}{0}
\setcounter{table}{0}
\makeatletter
\renewcommand{\theequation}{S\arabic{equation}}
\renewcommand{\thefigure}{S\arabic{figure}}
\renewcommand{\bibnumfmt}[1]{[S#1]}
\renewcommand{\citenumfont}[1]{S#1}

\subsection{Derivation of the scattering matrices}
\subsubsection{In-plane phonons}
In the harmonic approximation, the {\bf q}-representation of the elastic energy of in-plane deformation [Eq.~({\ref{energy_inplane}})] reads
\begin{equation}
\overline{E}_{\bf q} = \frac{1}{2}\sum_{\bf q}u^*_{q_{\alpha}}A_{\alpha\beta}^{\bf q}u_{q_{\beta}},
\end{equation}
where $A_{\alpha\beta}^{\bf q}$ is the force-constant matrix $A^{\bf q}_{\alpha\beta}=\frac{\partial^2 \overline{E}_{q}}{\partial u_{q^*_{\alpha}} \partial u_{q_{\beta}}}$ 
whose explicit form in case of orthorhombic symmetry is given by
\begin{equation}
(A^{\bf q}) =
\begin{pmatrix}
C_{11}q_x^2 + C_{66}q_y^2 & (C_{12}+C_{66})q_xq_y \\
(C_{12}+C_{66})q_xq_y & C_{22}q_y^2 + C_{66}q_x^2 \\
\end{pmatrix}.
\end{equation}
To evaluate $\langle |\overline{V}^{\mathrm{eff}}_{\bf q}|^2 \rangle$ [Eq.~(\ref{veffbar})], one needs to determine the inverse of $A^{\bf q}_{\alpha\beta}$, which for classical 
phonons ($\hbar\omega \ll k_BT$) corresponds to the correlation functions of the form $\langle u^*_{q_{\alpha}}u_{q_{\beta}} \rangle =k_BT(A^{\bf q})^{-1}_{\alpha\beta}$. 
Substituting $\langle u^*_{q_{\alpha}}u_{q_{\beta}} \rangle$ into Eq.~(\ref{veffbar}) one obtains the resulting expression for the scattering matrix of in-plane phonons 
$\langle |\overline{V}^{\mathrm{eff}}_{\bf q}|^2 \rangle$ given by Eq.~(\ref{Veff_inplane}).

\subsubsection{Flexural phonons}

General expression of the scattering probability due to the flexural phonons follows from the definition of $\widetilde{V}^{\mathrm{eff}}_{\bf q}$ given in the main text and has the form
\begin{eqnarray}
\langle |\widetilde{V}^{\mathrm{eff}}_{{\bf q}}|^2 \rangle=\widetilde{g}_{\alpha\beta}\widetilde{g}_{\mu\nu}\sum_{{\bf p}_1{\bf p}_2}{p}_{1\alpha}({q}_{\beta}-{p}_{1\beta}){p}_{2\mu}(-{q}_{\nu}-{p}_{2\nu})\nonumber \\
\times \langle    h_{{\bf p}_1} h_{{\bf q}-{\bf p}_1} h_{{\bf p}_2} h_{-{\bf q}-{\bf p}_2} \rangle.  \quad \quad \quad \quad \quad \quad
\label{veff2}
\end{eqnarray}
The expression in the angular brackets can be simplified by utilizing the Wick's theorem, which yields
$\langle ... \rangle = \delta_{{{\bf p}_1,{\bf q}+{\bf p}_2}} \langle h_{{\bf p}_1} h_{{-\bf p}_1} \rangle \langle h_{{\bf p}_1-{\bf q}} h_{{\bf q}-{\bf p}_1} \rangle $. Substituting
this expression into Eq.~({\ref{veff2}}) and neglecting off-diagonal elements of $\widetilde{g}_{\alpha\beta}$ as they are associated with shear deformations having significantly less effect on the 
electronic structure, we obtain a simplified expression given by Eq.~(\ref{Veff_integral}), which can be rewritten in the integral form as
\begin{widetext}
\begin{equation}
\langle |\widetilde{V}^{\mathrm{eff}}_{{\bf q}}|^2 = k_B^2T^2 \int_0^{\infty} \frac{\mathrm{d}^2{\bf p}}{(2\pi)^2}
\frac{[\widetilde{g}_x p_x(p_x-q_x)+\widetilde{g}_y p_y(p_y-q_y)]^2}{({\sqrt \kappa_x}p_x^2 + {\sqrt \kappa_y}p_y^2)^2({\sqrt \kappa_x}(p_x-q_x)^2 + {\sqrt \kappa_y}(p_y-q_y)^2)^2}
\end{equation}
\end{widetext}
where we have used the explicit form of the correlation functions $\langle h_{\bf q} h_{-\bf q}\rangle$ from Eq.~(\ref{corr_func}) and extended the integral over the Brillouin zone to the whole
space as it converges rapidly at large $|{\bf p}|$. To further simplify the expression, we make the transformation,
\begin{equation}
{\bf p} \rightarrow K^{-\frac{1}{4}}\left({\bf p'}+{\bf q'}\right) \text{ and } {\bf q} \rightarrow 2K^{-\frac{1}{4}}{\bf q'},
\end{equation}
where $K$ is a matrix with diagonal elements $\kappa_x$ and $\kappa_y$. After the transformation, we arrive at
\begin{equation}
\langle |\widetilde{V}^{\mathrm{eff}}_{{\bf q}}|^2 = k_B^2T^2 \int_0^{\infty} \frac{\mathrm{d}^2{\bf p'}}{(2\pi)^2}
\frac{   \left(a_x({p'}_x^2-{q'}_x^2) + a_y({p'}_y^2-{q'}_y^2 ) \right)^2   }            { |{\bf p'}+{\bf q}'|^4 |{\bf p'}-{\bf q}'|^4 }
\label{int3}
\end{equation}
where $a_x=\widetilde{g}_x/\sqrt{\kappa_x}$ and $a_y=\widetilde{g}_y/\sqrt{\kappa_y}$. Introducing polar coordinates for ${\bf p}'$ and evaluating the integral over the polar angle, we end up with an
integral which is logarithmically divergent at $|{\bf p}'|=|{\bf q}'|$.
The divergence is cut off at $|{\bf p}'|-|{\bf q}'|\sim|{\bf q}'^{*}|$, as for wave vectors ${\bf q}\lesssim{\bf q}^*$ the correlation function $\langle h_{\bf q} h_{-\bf q} \rangle \sim 1/|{\bf q}|^{4-\eta}$ with $\eta \approx 0.85$ \cite{Katsnelson-Book} instead of $\sim 1/|{\bf q}|^4$ [cf. Eq.~(\ref{corr_func})]. After taking the integral and some algebra, we obtain the resulting expression for $\langle |\widetilde{V}^{\mathrm{eff}}_{{\bf q}}|^2 \rangle$ given by Eq.~(\ref{Veff_flex}).

\subsection{Details of first-principles calculations}

DFT calculations of elastic constants and deformation potentials presented in Fig.~\ref{elastic}
were carried out by using the projected augmented-wave 
(PAW) formalism as implemented in the Vienna \emph{ab initio} simulation package 
({\sc VASP}) \cite{Blochl,Kresse1996a,Kresse1996b,Kresse1999}. To describe exchange-correlation
effects, we employed the gradient corrected approximation (GGA) in the parametrization of
Perdew-Burke-Ernzerhof (PBE) \cite{pbe}. An energy cutoff of 800 eV for the plane-waves and the
convergence threshold of 10$^{-10}$ eV were used in all cases. To avoid spurious interactions 
between the supercells, a vacuum slab of 20~\AA~thick was added in the direction perpendicular to
the sheet. Structural relaxation including the optimization of in-plane lattice parameters 
was performed until the forces acting on atoms were less than 10$^{-4}$ eV/\AA. 
In the unit-cell calculations of strain-energy curves a (48$\times$48) $\Gamma$-centered 
{\bf k}-point mesh was used to sample the Brillouin zone.

\begin{figure}[t]
\includegraphics[width=0.47\textwidth, angle=0]{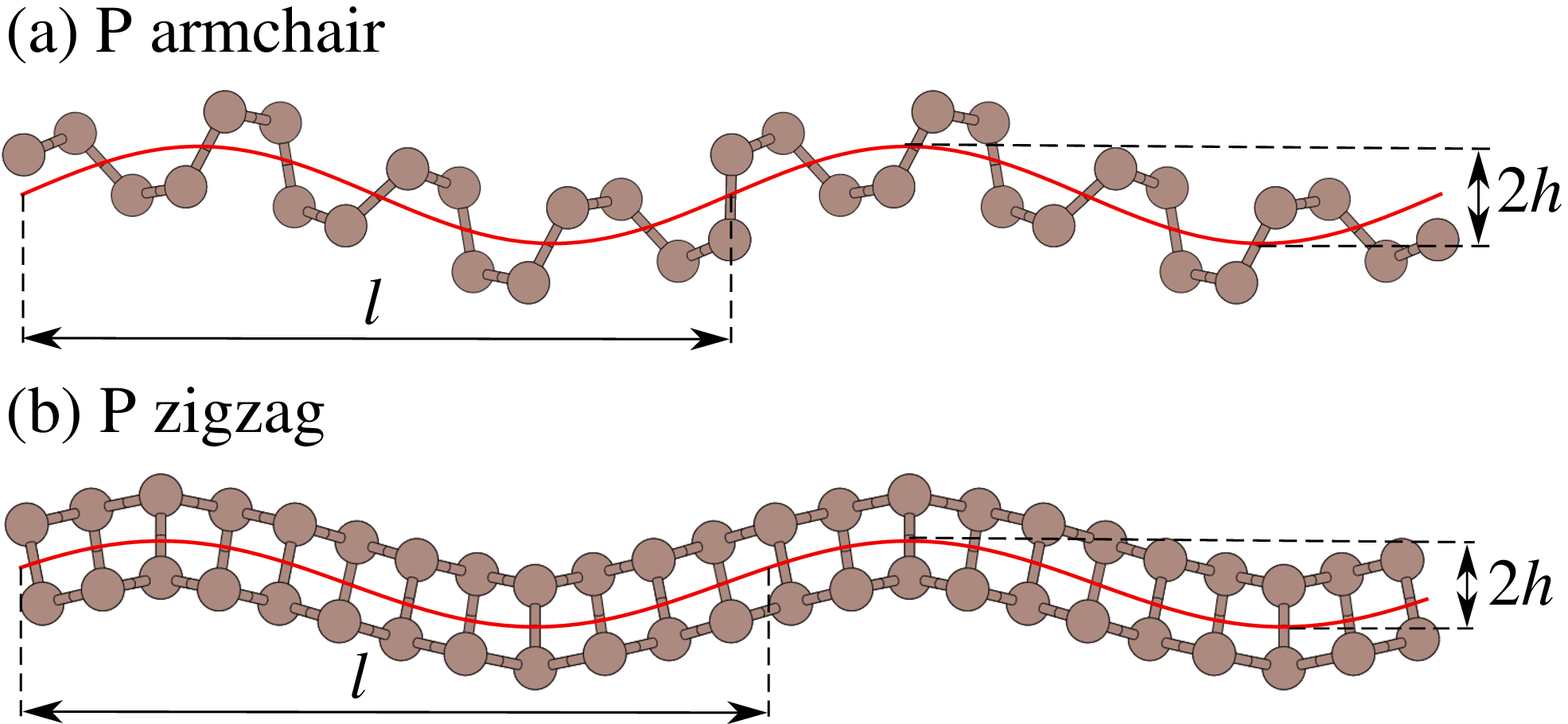}
\caption{(Color online) Schematic representation of a sinusoidal corrugation 
of BP used in this work to determine the flexural rigidity ($\kappa$)
 and deformation potential, corresponding to charge carrier interaction with 
flexural phonons ($\widetilde{g}$), from first-principles calculations. 
Red solid line shows the corrugations profile, whereas $h$ and $l$ determine 
its amplitude and period, respectively.}
\label{sine}
\end{figure}

To induce flexural deformations 
needed for the determination of flexural rigidities and deformation
potentials associated with flexural phonons, supercells with the dimensions of ($8a_x \times a_y$)
and ($a_x \times 11a_y$) were considered, respectively, for the armchair and zigzag directions of the 
deformation, where $a_x$ and $a_y$ are the unit cell vectors. 
The deformation is modeled by an 1D out-of-plane sinusoidal field
$h(x,y)=h \, \mathrm{sin}(qx)$ with $q=2\pi/l$, where $h$ and $l$ are the amplitude and period of the 
deformation, schematically shown in Fig.~\ref{sine}. In our case the deformation period is nearly
the same for both armchair and zigzag directions, which corresponds to $q\sim0.17$\AA$^{-1}$.
We then perform full structural relaxation for a series of fixed amplitudes $h$, which allows us to obtain
elastic energies as well as valence (or conduction) band edge shifts shown in Figs.~\ref{elastic}(b) and (d).
For corrugated supercells, a (4$\times$48) [or (48$\times$4)] {\bf k}-point mesh was used.

The electron and hole quasiparticle dispersion as well as the corresponding isoenergy contours presented in 
Fig.~\ref{dispersion} were obtained within the $GW_0$ approximation as described in Ref.~\onlinecite{Rudenko2015}.

\subsection{Energy-dependent effective mass approximation}

Dispersion of both electrons and holes in BP cannot be accurately described within
the constant effective mass approximation. Quasiparticle band structure calculated using $GW_0$ 
approximation exhibits deviations from the parabolic dispersion at low energies, 
which yields energy-dependent density of states (DOS) [Figs.~\ref{dispersion}(a) and (b)]. Here, to
describe the carrier dispersion $\varepsilon_{{\bf k}}$, we use the following implicit expression

\begin{figure}[!bp]
\includegraphics[width=0.50\textwidth, angle=0]{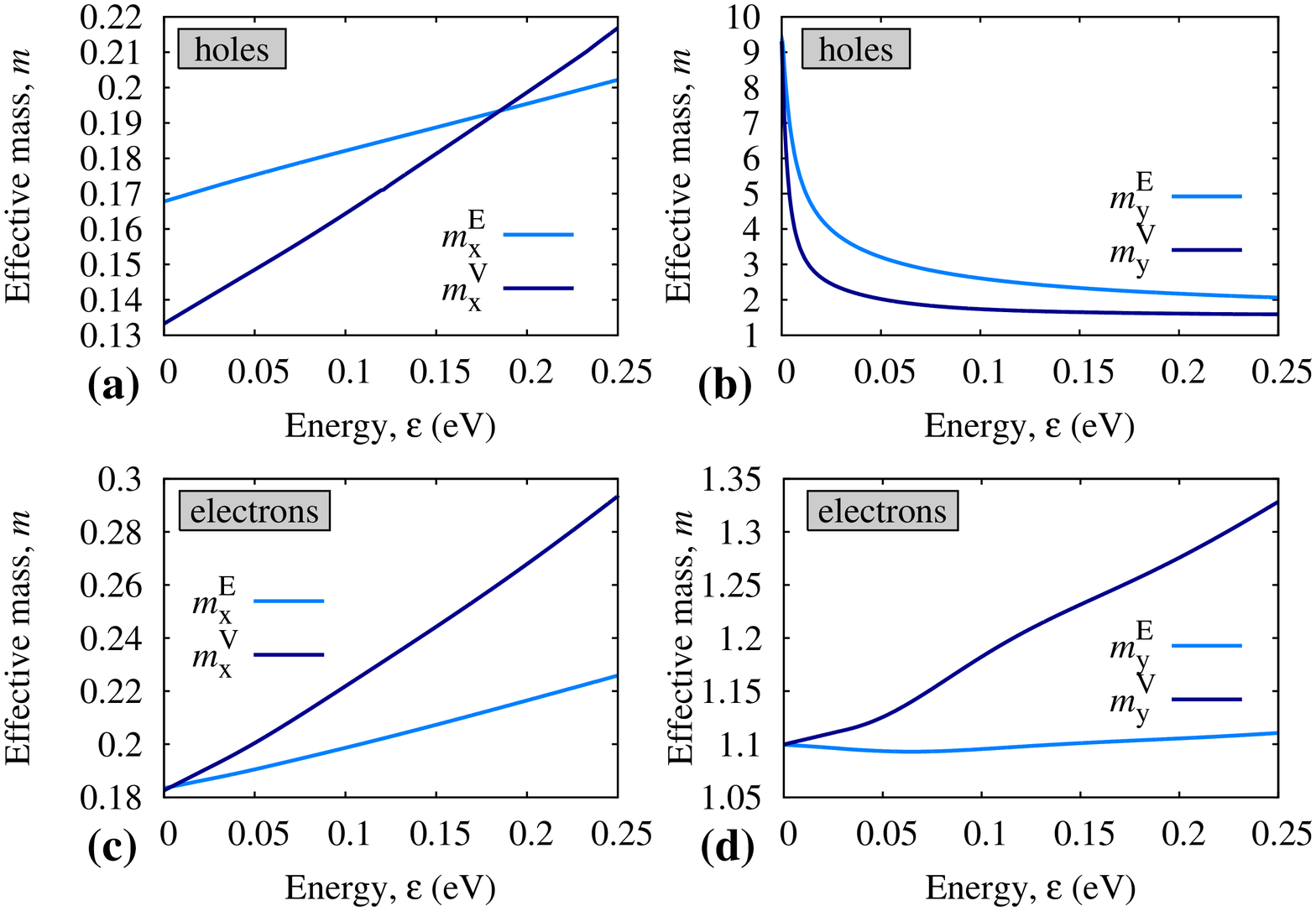}
\caption{(Color online) Energy dependence of the effective masses
used in this work to approximate anisotropic dispersion ($m^E_x$, $m^E_y$) 
and band velocities ($m^V_x$, $m^V_y$) related to holes and electrons in BP.}
\label{masses}
\end{figure}

\begin{equation}
\varepsilon_{{\bf k}}=\frac{\hbar^2k_x^2}{2m^E_x(\varepsilon)}+\frac{\hbar^2k_y^2}{2m^E_y(\varepsilon)},
\label{disp}
\end{equation}
where parameters $m^E_x(\varepsilon)$ and $m^E_y(\varepsilon)$ are energy-dependent effective masses corresponding to the armchair and zigzag directions, 
respectively.
For slowly varying masses, the corresponding DOS can be approximated as $N(\varepsilon)=(g_s/2\pi\hbar^2)[m^E_x(\varepsilon)m^E_y(\varepsilon)]^{1/2}$, with $g_s=2$ being the spin-degeneracy factor.
Here, we determine the masses from the normalization of $N(\varepsilon)$ to the number of particles and from the condition
$m^E_y(\varepsilon)/m^E_x(\varepsilon)=\mathrm{max}[{\bf k}(\varepsilon)]/\mathrm{min}[{\bf k}(\varepsilon)]$, where ${\bf k}(\varepsilon)$ is the quasiparticle wave vector at
the energy $\varepsilon$. The validity of the model given by Eq.~(\ref{disp}) follows from Fig.~\ref{dispersion}, which shows good agreement with the $GW_0$ results for 
the dispersion and isoenergy contours. To describe band velocities $v^{x(y)}_{\bf k}$, we follow similar scheme, 
\begin{equation}
v^{x(y)}_{{\bf k}}=\frac{\hbar k_{x(y)}}{m^V_{x(y)}(\varepsilon)},
\label{velo}
\end{equation}
but use another set of energy-dependent parameters $m^V_x(\varepsilon),m^V_y(\varepsilon)$ which are chosen such that $m^V_y(\varepsilon)/m^V_x(\varepsilon)=m^E_y(\varepsilon)/m^E_x(\varepsilon)$ and 
$\langle (\partial \varepsilon_{\bf k}/\partial k_{x(y)})^2 \rangle_{\varepsilon} = \varepsilon/m^V_{x(y)}$, where $\langle ... \rangle_{\varepsilon}$ means averaging over the isoenergy contour with 
the energy $\varepsilon$.

In Fig.~\ref{masses}, we show energy-dependence of effective masses 
$m^E_x,m^E_y$ and $m^V_x,m^V_y$ used to describe the dispersion
$\varepsilon_{\bf k}$ [Eq.~(\ref{disp})] and band velocities 
$v^{x(y)}_{\bf k}$ [Eq.~(\ref{velo})] of electron and holes in BP.

\subsection{Two-phonon contribution to the carrier mobility}

\begin{figure}[!tbp]
\includegraphics[width=0.48\textwidth, angle=0]{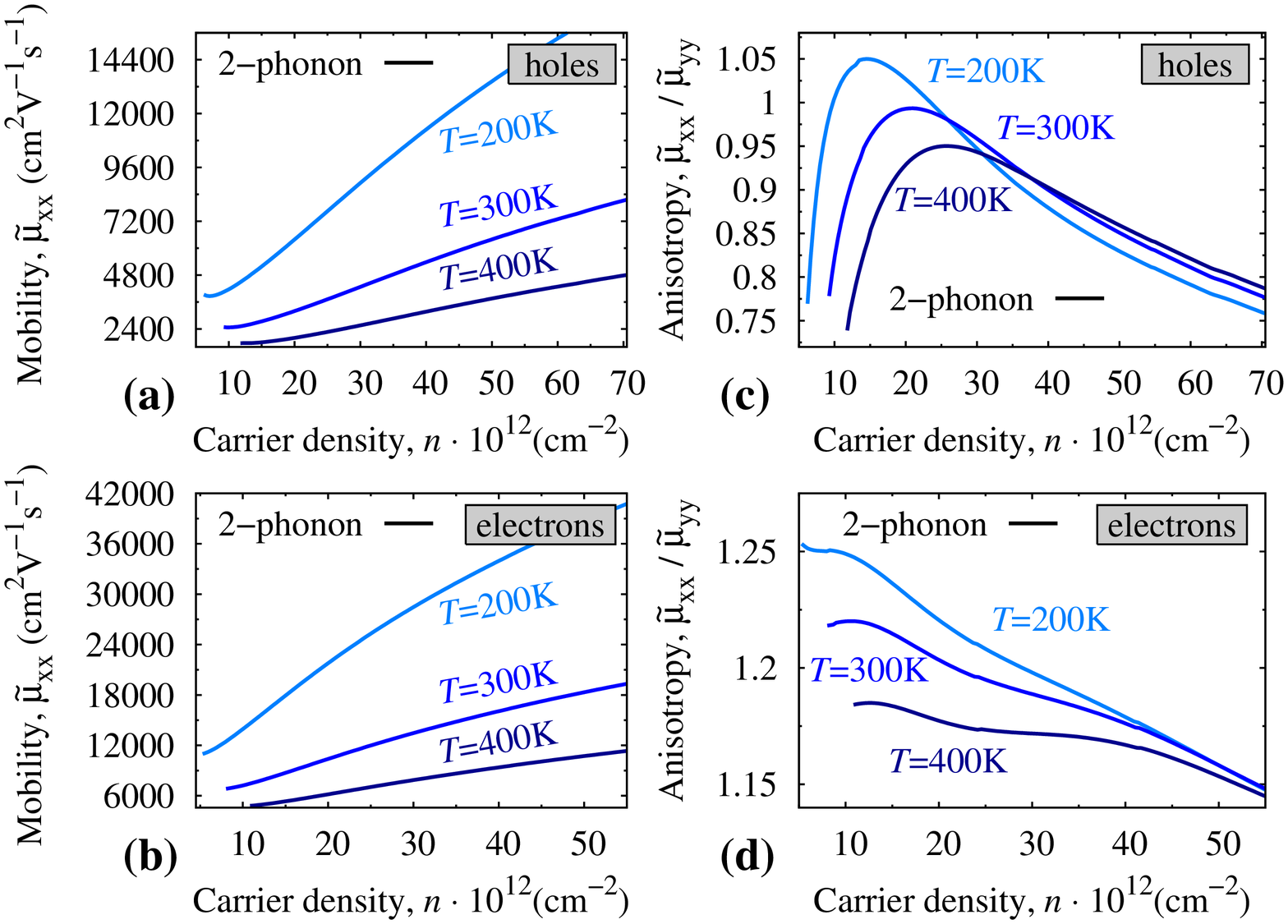}
\caption{(Color online) (a,b) Contribution to the intrinsic mobility
of BP related to two-phonon processes ($\widetilde{\mu}_{xx}$) shown as a function of
the carrier concentration ($n$) calculated along the armchair direction
for different temperatures ($T$). 
(c,d) Anisotropy of the two-phonon contribution to the carrier mobility 
represented at a ratio between the 
mobilities along the armchair and zigzag direction ($\widetilde{\mu}_{xx}/\widetilde{\mu}_{yy}$) shown
for different $T$. The lowest
depicted density corresponds to the regime with $\mathrm{ln}(\overline{k}/q^*)>1$.}
\label{mobility_flex}
\end{figure}

In Fig.~\ref{mobility_flex}, we show the contribution to the mobility associated with
two-phonon processes ($\widetilde{\mu}_{xx}$) due to scattering by flexural phonons.

\subsection{Carrier mobility within the constant effective mass approximation}

The integrals determining the conductivity given by Eqs.~(\ref{sigmax}) and (\ref{taux})
can be evaluated analytically if purely parabolic dispersion of charge carriers is
assumed,

\begin{equation}
\varepsilon_{{\bf k}}=\frac{\hbar^2k_x^2}{2m_x}+\frac{\hbar^2k_y^2}{2m_y}.
\label{disp2}
\end{equation}

The corresponding Fermi contour is now represented by an ellipse determined by constant effective masses $m_x$ and $m_y$, resulting in
an energy-independent density of states, $N(\varepsilon)=g_s\sqrt{m_xm_y}/2\pi\hbar^2$. Assuming that the electron gas is degenerate ($\varepsilon_F \gg k_BT$), 
the $x$-component of the mobility within this approximation is given by

\begin{equation}
\mu_{xx} = \frac{e\tau_{xx}}{2m_x},
\end{equation}
whereas $\tau_{xx}$ takes the form

\begin{equation}
\tau^{-1}_{xx} = \frac{\pi}{\hbar N(\varepsilon )\langle v^{2}_x \rangle_\varepsilon}\sum_{{\bf k}{\bf k}'}\delta(\varepsilon_{\bf k}-\varepsilon_F) \delta(\varepsilon_{\bf k'}-\varepsilon_F)v^{x2}_{\bf q} \langle |V^{\mathrm{eff}}_{\bf q}|^2 \rangle,
\label{taux_approx}
\end{equation}
where ${\bf q}={\bf k}-{\bf k'}$. Substituting Eq.~(\ref{Veff_inplane}) into (\ref{taux_approx}) and noting that $2C_{66} \approx \sqrt{C_{11}C_{22}}-C_{12}$ which 
represents a reasonable approximation for BP, we obtain the following scattering rate for single-phonon processes

\begin{equation}
\overline{\tau}^{-1}_{xx} =\frac{k_BT\overline{g}^2_xm^*}{\hbar^3C_{11}} \overline{A}_{xx}
\end{equation}
where $m^*=\sqrt{m_xm_y}$ and $\overline{A}_{xx}$ can be regarded as a single-phonon anisotropic factor,

\begin{align}
\overline{A}_{xx} = \frac{1}{(1+\sqrt{s})^3 \sqrt{s}} \left\{ s^{1/2} + 3s + s^{3/2} \right. \quad \quad \quad \quad \quad \nonumber \\ 
\left. + s^{1/2}\left[2t + \frac{m_y}{C_{66} m_x}\left(\sqrt{C_{22}}-\sqrt{C_{11}}\frac{\overline{g}_y}{\overline{g}_x}\right)^2 \right] + t^2 \right\},
\end{align}
where we used the notation $s=\sqrt{\frac{C_{22}}{C_{11}}}\frac{m_y}{m_x}$, and $t=\frac{\overline{g}_ym_y}{\overline{g}_xm_x}$. As can be easily checked, without anisotropy $\overline{A}_{xx}$ is unity. 

The scattering rate for two-phonon processes is obtained similarly using Eqs.~(\ref{Veff_flex}) and (\ref{taux_approx}). The resulting expression reads

\begin{equation}
\widetilde{\tau}^{-1}_{xx} = \frac{k^2_BT^2\widetilde{g}^2_x}{8\pi\hbar\varepsilon_F\kappa_x^2} \mathrm{ln}(\sqrt{\mathrm{e}}\gamma) \widetilde{A}_{xx},
\end{equation}
where $\widetilde{A}_{xx}$ is now a two-phonon anisotropic factor,

\begin{align}
\widetilde{A}_{xx} = \frac{l^{1/2}r^{-1}}{\mathrm{ln}(\sqrt{\mathrm{e}}\gamma)} \left\{ \frac{(1-p)^2}{2(1+\sqrt{l})}+\frac{(1+p)^2}{4(1+\sqrt{l})} + \frac{(1+p^2)}{(1+\sqrt{l})}\mathrm{ln}\gamma \right . \nonumber \\
\left. - \frac{(3\sqrt{l}+l)}{(1+\sqrt{l})^3} \frac{(1-p)^2}{2} + \frac{(1-p^2)}{(1+\sqrt{l})^2}\mathrm{ln}\gamma \right\},
\end{align}
with $r=(\kappa_y/\kappa_x)^{1/2}$, $p=\frac{\widetilde{g}_y/\widetilde{g}_x}{(\kappa_y/\kappa_x)^{1/2}}$, and $l=rm_y/m_x$.


\begin{thebibliography}{99}
\bibitem{Overlap}
Our first-principles estimates show that at carrier density $n=10^{13}$ cm$^{-2}$ the contribution of the wave function overlap to the scattering matrix does not exceed a few percent. For example, for holes $|\langle -k^x_F | k^x_F \rangle|^2 = |\langle -k^y_F | k^y_F \rangle|^2 = 0.94$, and $|\langle k^x_F | k^y_F \rangle|^2 = 0.96$.
\bibitem{Remark}
We use elastic parameters for graphene from Ref.~[\onlinecite{Kudin}] and consider the logarithmic term $\mathrm{ln({\sqrt e}\gamma)}\approx 2.5$ to be the same as in BP. 
The corresponding density of states reads $N(\varepsilon_F)=(2 / \pi\hbar^2) (\varepsilon_F / v_F^2)$, where $v_F\approx10^6$ m/s is the Fermi velocity in graphene.
\end{thebibliography}
\end{document}